\documentclass[12pt,preprint]{emulateapj}

\newcommand{\mymail}{oysteol@astro.uio.no}
\begin{document}

\altaffiltext{1}{also at Center of Mathematics for Applications, University of 
Oslo, P.O. Box 1053 Blindern, N-0316 Oslo, Norway}

\title{Spectroscopic measurements of dynamic fibrils 
in the Ca~{\small{II}}~8662~{\AA} line.}

\author{{\O}ystein Langangen}
\author{Mats Carlsson \altaffilmark{1}}
\author{Luc Rouppe van der Voort \altaffilmark{1}}
\author{Viggo Hansteen \altaffilmark{1}}
\affil{Institute of Theoretical Astrophysics, University of Oslo, 
P.O. Box 1029 Blindern, N-0315 Oslo, Norway}
\email{\mymail}
\and
\author{Bart De Pontieu}
\affil{Lockheed Martin Solar and Astrophysics Lab, 3251 Hanover St., 
Org. ADBS, Bldg. 252, Palo Alto, CA 94304, USA}

\begin{abstract} 
We  present high spatial resolution spectroscopic measurements of 
dynamic fibrils (DFs) in the Ca~{\small{II}}~8662~{\AA} line. 
These data show clear Doppler shifts in the identified DFs, 
which demonstrates that at least a subset of 
DFs are actual mass motions in the chromosphere. 
A statistical analysis of 26 DFs reveals a strong and statistically 
significant correlation between the maximal velocity and the deceleration.
The range of the velocities and the decelerations are substantially lower,
about a factor two, in our spectroscopic observations compared to the 
earlier results based on proper motion in narrow band images.
There are fundamental differences in the different observational 
methods; when DFs are observed spectroscopically the measured Doppler shifts
are a result of the atmospheric velocity, weighted with the 
response function to velocity over an extended height. 
When the proper motion of DFs is observed 
in narrow band images, the movement of the top of the DF is observed.
This point is sharply defined because of the high contrast between the 
DF and the surroundings. The observational differences between
the two methods are examined by
several numerical experiments using both numerical simulations and 
a time series of narrow band H$\alpha$ images.
With basis in the simulations we conclude that the lower maximal velocity is
explained by the low formation height of the Ca~IR~line.
We conclude that the present observations support the earlier result that DFs 
are driven by magneto-acoustic shocks exited by convective flows and p-modes. 
\end{abstract}


\section{Introduction}\label{intro}
The dynamical nature of the chromosphere is obvious when the Sun
is imaged in the line center of strong chromospheric spectral lines,
most commonly in the H$\alpha$ line \citep[e.g.,][]{2006Noort}.  
One of the dominating features is a vast number of thin (0.2-1 Mm) 
omnipresent jet like structures \citep{1968Beckers}.
On the quiet solar limb they are commonly known as spicules, while on the
quiet disk they are often called mottles, and finally in active regions 
they are known as active region fibrils or dynamic fibrils (DFs).
The nomenclature can be confusing, but there are strong indications that these 
structures are physically closely related 
\citep{1994Tsiropoula,1995Suematsu,2001Christo,2007Rouppe}.

Recently, a combination of high--resolution observations and advanced 
numerical modeling have shown that DFs are most likely driven by shocks
that form when photospheric oscillations leak into the chromosphere
along inclined flux tubes
\citep{1990Suematsu,2004dePontieu,2006Hansteen,2007dePontieu}.
The inclination of the magnetic field lowers the acoustic cutoff frequency 
sufficiently to allow p--modes with the dominant low frequencies to propagate
along flux tubes \citep{1973Mich,1977Bel}.

These insights into the formation of DFs have become possible because 
of recent developments in observational techniques, such as
bigger telescopes combined with real time wavefront 
corrections by adaptive optics (AO) systems \citep[e.g.][]{2000Rimmele,SSTAO}, 
and post-processing methods \citep[e.g.][]{speckle,momfbd}
which have made observations of these jet structures much more reliable.

These developments have spurred several authors to focus on the detailed 
understanding of DFs \citep[e.g.,][]{2003dePontieu,2004dePontieu,
2005dePontieu,2007dePontieu,2004Kostas,2006Hansteen,2006deWijn,2007Julius,
2007Lars}. 
One of the important results of this work is that the DFs are 
driven by and can channel photospheric oscillations into the 
chromosphere and the corona. 

\begin{figure*}[!ht]
\includegraphics[width=\textwidth]{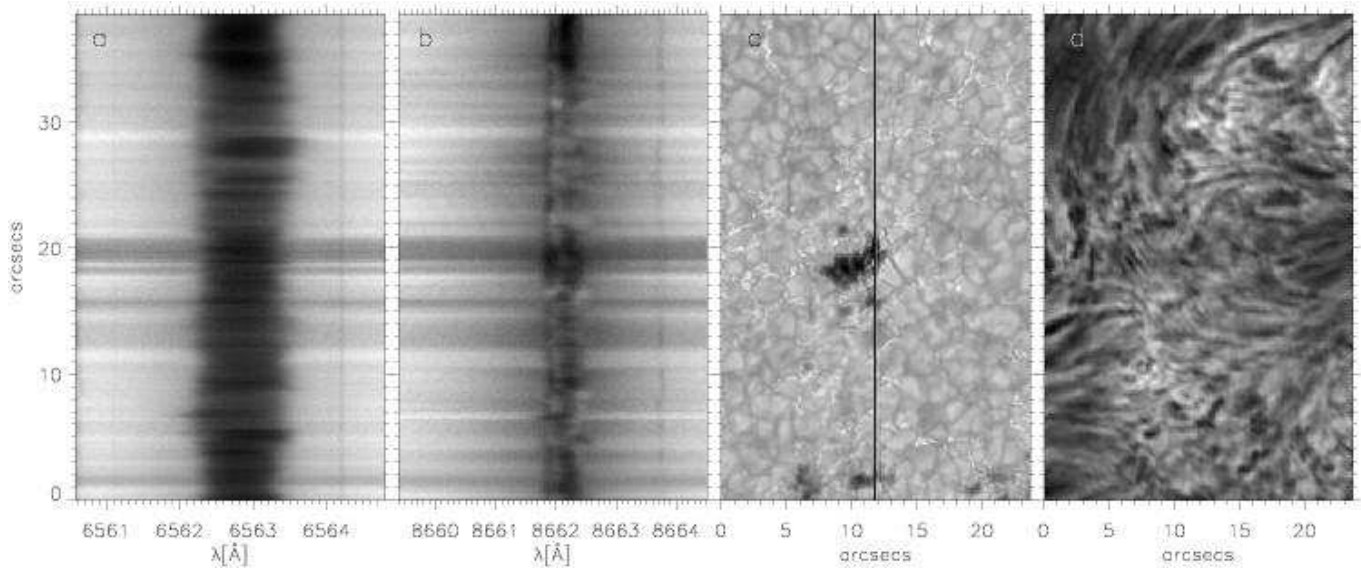}
\caption{
Spectrograms of the H$\alpha$ line (panel a) and 
the Ca~{\small{II}}~8662~{\AA} line (panel b). Notice the
highly dynamical line center of the Ca line.
The corresponding MFBD processed slitjaw image (panel c)
and the narrow band H$\alpha$ DOT image (panel d). The position of 
the spectrograph's slit is marked with a line in the DOT image.  
}
\label{plotone}
\end{figure*}

In two papers \citet{2006Hansteen} and \citet{2007dePontieu} used high 
spatial and high cadence observations of the H$\alpha$ line center 
together with realistic simulations to investigate the nature of DFs. 
One of their conclusions was that the DFs follow parabolic paths along 
their axis with decelerations lower than the solar gravitational
deceleration. Previous observations did not allow an accurate determination 
of the nature of the trajectory because of lower quality
data and line--of--sight (LOS) effects that were difficult to estimate
\citep{1988Nishikawa,1995Suematsu}. 

\citet{2006Hansteen} and \citet{2007dePontieu}
further report regional differences between DFs 
observed in two different plage areas. The regional differences are 
explained by different inclination angles of the magnetic fields in the 
two regions. In this way the magnetic topology of the solar atmosphere works 
as a filter, where only waves with certain periods can leak through. 
The simulations, spanning from the upper convection zone to the corona, 
reproduce the observed correlations between the maximum velocities and 
decelerations in DFs leading to the conclusion that DFs are formed by 
chromospheric shocks driven by global p-modes and convective flows.

In this paper we add to the understanding of the DFs, by analyzing 
high spatial and high cadence spectrograms of the 
Ca {\small{II}} 8662 {\AA} line, put into context by simultaneous H$\alpha$
spectrograms and narrow-band images. Furthermore, the observational 
results are compared with the numerical simulations of \citet{2006Hansteen}.

In \S~\ref{Obs} we describe the observing program and 
instrumentation. The data reduction method is described in \S~\ref{Data}.
In \S~\ref{Obsres} we show the results of the observations.
The main observational errors are discussed in \S~\ref{error}.
To get a better understanding of the observations we present several 
numerical experiments in \S~\ref{Sim}. Finally, we summarize the results 
in \S~\ref{Con}.

\section{Observing program and instrumentation}\label{Obs}

The data presented in this paper were obtained in a co-observation 
campaign between the Swedish 1-m Solar Telescope (SST) \citep{SST} and 
the Dutch Open Telescope (DOT) \citep{DOT}. 
The spectrograms were obtained with the SST, using the TRIPPEL 
spectrograph\footnote{\scriptsize{
http://dubshen.astro.su.se/wiki/index.php/TRIPPEL$\_$spectrograph}}
in combination with the adaptive optics system at the SST \citep{SSTAO}.
In short the TRIPPEL spectrograph is a multi port spectrograph, using 
an echelle grating with a blaze angle of $63.43^\circ$.
The theoretical spectral resolution is 240000. 
In our case we observe in two ports to obtain simultaneous
H$\alpha$ and Ca~{\small{II}} 8662 {\AA} spectra. We observe at $-1.06^\circ$
away from the blaze angle in order 34 and 26 respectively. 
To remove light from other orders we 
use two standard filters centered at 6562 and 8680 with FWHM of 43 
and 101 {\AA} respectively. These filters are placed in front of two Megaplus
1.6 cameras (KAF--0401 Image sensor). With this setup the spatial 
pixel size is $0.0411\arcsec{}$\,pixel${}^{-1}$ for both spectral regions 
while the spectral pixel size is $0.0105$\,{\AA}\,pixel${}^{-1}$ and 
$0.0129$\,{\AA}\,pixel${}^{-1}$ for the H$\alpha$-- and Ca {\small{II}} IR --
spectral regions respectively. 
To obtain a reasonable signal--to--noise ratio we observe with 80~ms 
exposure times with both cameras.
For each of the spectrogram cameras a slit-jaw camera was operated 
with the same exposure time and with a wideband filter centered close
to the corresponding spectrogram wavelength. The filters used
were centered at $6565$~{\AA} and $8714$~{\AA} and had FWHM of
$10$~{\AA} and $100$~{\AA} respectively. These slit-jaw images were used 
for the destretching of the spectrograms, see \S~\ref{Data}.
Furthermore, a slit-jaw camera was operated with an exposure time
of $8$~ms to obtain reference slit-jaw images less blurred by seening.
The filter used was centered at $6565$~{\AA} with a FWHM of $10$~{\AA}.
To get context images for the spectrograms we obtain
co-temporal and co-spatial observations of both the H$\alpha$ line center 
and H$\alpha$ continuum using the DOT. 

On 2006 May 04 we used a small pore in the 
NOAA active region 10878 (N14, E04, $\mu=0.96$) as adaptive optics lock point
for the SST, see Fig.1, to obtain a time series of about 40 minutes 
(08:20:33--09:01:14 UT) during good to excellent seeing conditions. During
this time period we observe in ``save all'' mode, which in our case means 
a cadence of $\sim0.5$ seconds. 
All data presented in this paper are from this time series. An overview of the
observations can be seen in Fig.\ref{plotone}.

\section{Data reduction}\label{Data}

\subsection{Aligning, destretching, and  Fourier filtering}\label{align}
Flat field and dark current images were constructed from the mean of
500 images, following the same procedure as described in
\citep{2007Langangen}. After these corrections, the spectrograms 
are aligned to the corresponding wide band slit-jaw images.
This is done by correlating the continuum intensity in the spectrogram
and the intensity close to the slit in the slit-jaw image.
Furthermore, the continuum images from the DOT are aligned with the slit-jaw
images, and since the H$\alpha$ narrow band images on the DOT are aligned with
the continuum images we get a series of aligned spectrograms and 
both continuum and narrow band images, as seen in Fig.\ref{plotone}.

Since the AO is not able to compensate for all the modes in the seeing, 
especially since the exposure times are 80 ms, post processing of
the spectrograms as well as the images is desirable.
Both the narrow-band and wide-band DOT images are 
post--processed using the speckle method \citep{speckle}. 
The slit--jaw images obtained with $8$ ms exposure times are post--processed 
using the multi frame blind deconvolution \citep[MFBD,][]{momfbd}.
Due to the lack of spectral information outside the slit of the 
spectrograph, a similar reconstruction of the spectrograms is not possible. 
A partial one dimensional compensation is, however, possible. 
The amount of destretching is obtained from the slit-jaw image with the
same exposure time as the spectrogram. For this purpose we use the routines 
developed by \citet{1994DS}. The amount of destretching from the tile
covering the slit is then used to destretch the spectrogram. 
Due to the one dimensionality of this procedure it is clearly not a 
complete restoration method, hence the improvements are quite small. 

The spectrograms have a signal--to--noise ratio of typically 180 at continuum 
level. To reduce the noise level we apply a conservative 
low pass Fourier filter technique.
This reduces the noise significantly (about a factor two), 
without losing any significant signal.

\subsection{Wavelength calibration}\label{wavelength}

A mean solar spectrogram is constructed by 
adding the flat field spectra, altogether 500 exposures. 
The mean aberration corrected spectrum is then compared to the FTS atlas of 
\citet{1987Brault}. This atlas has proved to be well calibrated in wavelength
and it shows no systematic offset in line shifts with wavelength 
\citep{1998Allende}. Since the solar atlas is corrected for the Earth's 
rotation, the Earth's orbital motion and the Sun's rotation we automatically 
get a corrected spectrogram when we calibrate using the FTS atlas. 
Note that in this way our wavelength scale is the same 
as that of the FTS atlas; we therefore have to correct for the gravitational
redshift and the convective blueshift. The convective 'blueshift' is in fact
a redshift not caused by classical convection in the case of 
the Ca~{\small{II}}~IR lines \citep{2006Uitenbroek}.
We use the rest wavelength given by the VALD database 
\citep{1995Piskunov,vald,1999Ryabchikova}. 
We correct for the gravitational redshift of $634$~m\,s${}^{-1}$.
The residual redshift in the mean profile of $425$~m\,s${}^{-1}$ 
is caused by the intensity and velocity variations in the chromosphere,
and possibly also by errors in the rest wavelength \citep{2006Uitenbroek}. 
No correction of this residual redshift is done.

The slit covers a network region of about 38\arcsec{} or approximately 28 
Mm on the Sun. We have tested the difference between the ``quiet'' solar atlas
and our ``active region'' mean spectrum . Even though the ``active region'' 
mean spectrum is quite different with respect to the shape of the line 
profile, the mean line center shift differs only with about $70$~m\,s${}^{-1}$.
Since time dependent variations in the wavelength calibration are known to
be quite small we do not use any time dependent wavelength calibration.
All in all we estimate the wavelength calibrations to have an accuracy of
about $100$~m\,s${}^{-1}$, the errors due to uncertainties
in the rest wavelength are, however, of several hundred m\,s${}^{-1}$.

\begin{figure}[!ht]
\includegraphics[width=0.5\textwidth]{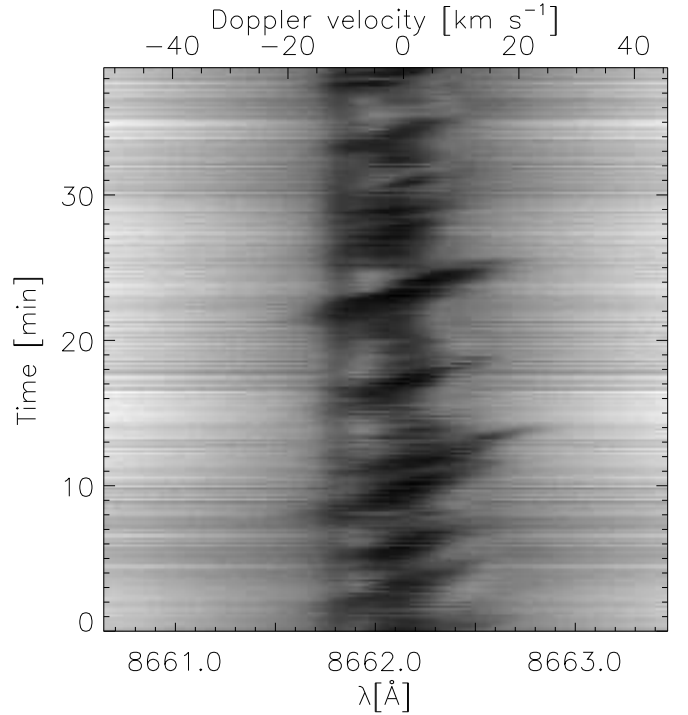}
\caption{Velocity-time plot for the Ca~II~8662~{\AA} line.
Several DFs are seen as diagonal dark components across the spectral line.
}
\label{plottwo}
\end{figure}
\begin{figure}[!ht]
\includegraphics[width=0.48\textwidth]{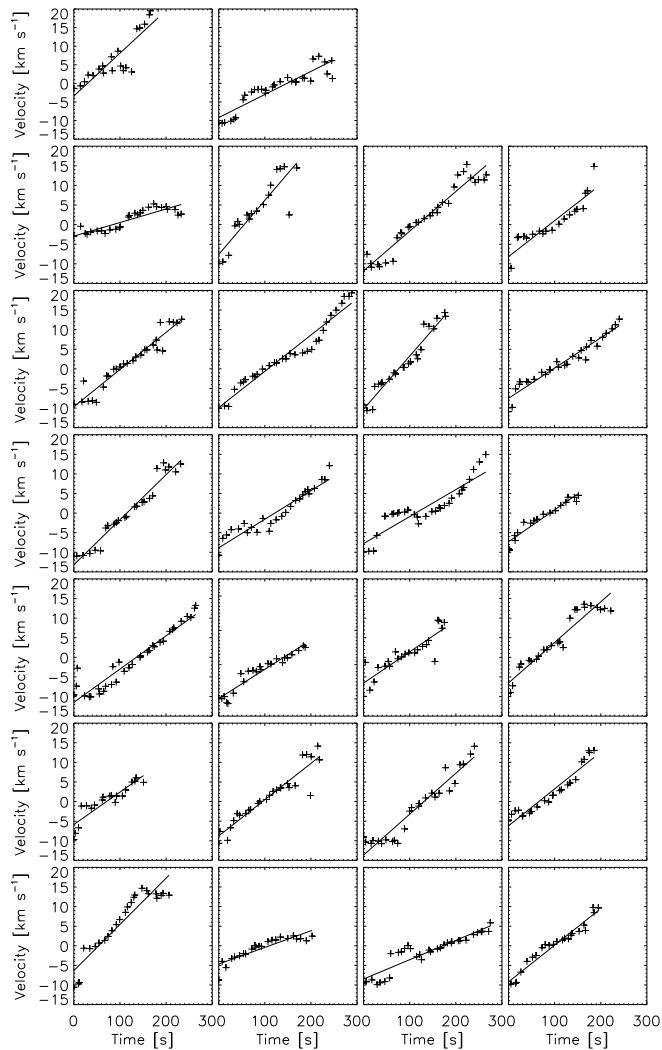}
\caption{Doppler velocities measured in 26 observed DFs (crosses). 
The least square linear fit to these points is plotted as a 
solid line in each panel.}
\label{plotthree}
\end{figure}

\begin{figure}[!ht]
\includegraphics[width=0.48\textwidth]{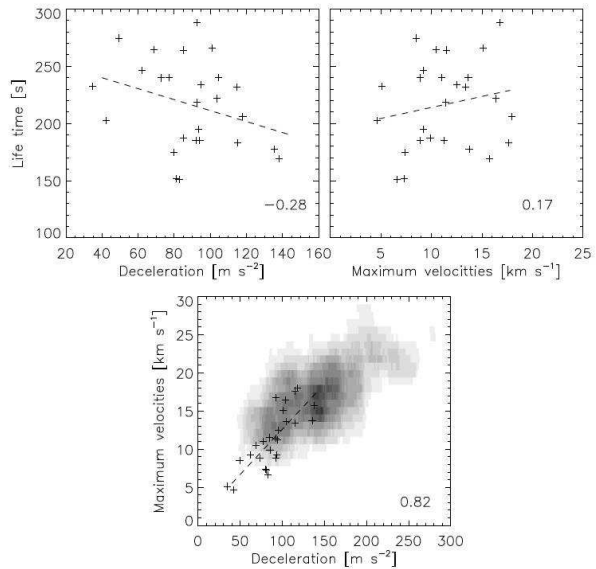}
\caption{Scatter plots of different observables.
The linear least square fit to the data is plotted as a dashed line. 
The Spearman correlation coefficients for each scatter plot 
are shown in the lower right corner of each panel.
The data obtained by \citet{2006Hansteen} and \citet{2007dePontieu} are
shown as an inverse gray--scaled density image in the lower panel.}
\label{plotfour}
\end{figure}

\section{Observational results}\label{Obsres}

Since the H$\alpha$ line is a very wide spectral line with a flat center, 
it is not very sensitive to Doppler shifts. In addition, the difficulties 
of understanding the line formation, make H$\alpha$ less than 
ideal for measuring Doppler shifts and interpreting the results. Instead, 
we use the Ca~{\small{II}}~8662~{\AA} line to measure Doppler 
shifts, only using the H$\alpha$ line as a reference when we identify DFs.
The reader should note that the Ca~{\small{II}}~8662~{\AA} line
also has several disadvantages. Firstly, it has a Fe blend in the blue wing, 
at $8661.9$~{\AA}, that
can interfere with large blueshifts in the line center. Secondly the line 
is very dynamic and shows several absorption and emission components, which 
complicates Doppler measurements. This line is nevertheless suited to
measure the Doppler shifts in the DFs, both because it is quite narrow with
steep wings and because DFs are easily identified as dark components moving 
across the line, as seen in Fig.\ref{plottwo}. 

Since the spectrograms suffer from residual seeing effects, the 
solar features (e.g. DFs, bright points etc.) will move
somewhat both across and along the spectrograph slit. 
To reduce the effect of this kind of movement we re-bin the spectrograms 
to a spatial pixel-size of $0.12\arcsec{}$, and use only the highest contrast
images in 6 seconds bins for this analysis.

To visualize the DFs in a simple way and to make the identification
of DFs easier, we make velocity-time (VT) plots, see Fig.\ref{plottwo}.
The DFs are identified by the diagonal dark components seen in the
VT plots. To avoid miss-identifications of DFs we only identify DFs above 
the plage region in the lower part of the continuum images. 
The narrow band images show that this region 
clearly contains DFs, but the spatial resolution of the narrow-band images 
is not good enough to measure the motion of independent DFs. Despite the 
lack in spatial resolution we look for a drop of the intensity 
in the narrow-band images to accept a DF identification. Furthermore, we 
look for a co--spatial and co--temporal component in the H$\alpha$ spectrograms.
In this way we manually identify 26 DFs.

We measure the Doppler shift in these DFs by fitting a 4th order 
polynomial to the dark component, 
first appearing in the blue wing, then tracking it through the line center,
before it appears in the red wing. 
The identification of the components is also done manually.
The resulting Doppler measurements of the DFs are shown in Fig.\ref{plotthree}.

A linear polynomial is fitted to each of the observed DFs.
Tests with the reduced $\chi^2$--method show that this 
fit is a good fit if the error is estimated to about 
$2$~km\,s${}^{-1}$. An error of $2$~km\,s${}^{-1}$ is reasonable for these
observations.

Accepting the linear fits we can derive a mean value for the 
deceleration of $89~{\pm}~25$~m\,s${}^{-2}$, for the maximum velocity of 
$11.3~{\pm}~3.8$~km\,s${}^{-1}$, and life times of $217~{\pm}~39$~secs,
with errors given in one standard deviation.
These values are all on the low side of earlier observations,
the reason for this discrepancy will be discussed in detail in \S\ref{Sim}.
It should also be mentioned that the maximum velocity in the receding phase
of the DF life, seems to have higher values than in the emerging phase,
on average $2.4$~km\,s${}^{-1}$ higher. This result may be caused by 
the Fe blend in the blue wing of the Ca line. 

We get a good correlation between the maximum velocity and the deceleration, 
see Fig.\ref{plotfour} lower panel. Using a Spearman rank correlation test
\citep[see e.g.][]{Numerical}
we get a correlation coefficient of $0.82$. With the null hypothesis of 
no correlation we get a significance of $4.1\sigma$ for falsification, 
which is a fairly strong statistical significance. This correlation 
has been shown to be a signature of the DF shock theory 
\citep{2006Hansteen,2007dePontieu,2007Lars}.
The weak negative correlation between the lifetime and the 
deceleration, see Fig.\ref{plotfour} upper left panel and the weak 
positive correlation between the maximum velocity and the lifetime, 
seen in the upper right panel of Fig.\ref{plotfour} have a statistical weak
significance of only $1.4\sigma$ and $0.9\sigma$ respectively. 
Similar weak correlations were found in \citet{2007dePontieu}.

\section{Error analysis}\label{error}

The observations presented in this paper differ from previously
presented observations of DFs in some fundamental ways. Our data contain the 
full spectroscopic information, while previous results have usually been 
based on narrow band images from one or a small number of spectral positions.
This makes our observations more robust when it comes to Doppler velocity 
measurements compared to other methods. Unfortunately this advantage is 
only achieved at the expense of other disadvantages. In our case there is
one main concern, namely the lack of spectral information from regions 
outside the slit. This lack of information leads to two problems in our
analysis. Firstly parts of the DFs are missed, thus altering the 
correlations between the observables.
Secondly it makes us unable to post-process the observations in a 
fully consistent way. Nevertheless these observations contain important
information of the DF dynamics. Since we can not correct for these 
shortcomings we will try to understand their effects on the observations.

To get an estimate of the magnitude and effect on the observables due
to the lack of information outside the slit, we investigate an
H$\alpha$ line-core imaging times-series, similar to the one presented
by \citet{2007dePontieu}. We pick a random ``slit'' covering a
similar plage region as in our spectroscopic observations.
We identify 11 DFs crossing the slit, and measure the movement of these
DFs. The part of the DFs outside the slit is removed to investigate the
effect of the 1D slit. Both the whole DF statistics and the reduced DF
statistics show the same correlation between the deceleration and maximum
velocity, but the correlation becomes weaker when parts of the DFs
are removed. This indicates that the spectroscopic results are still valid,
but somewhat affected. The mean values are typically reduced with about
$20\%$. We do expect this effect to be much less severe in 
Doppler measurements, since we are biased toward DFs that have strong LOS
components, and hence are smaller in the image plane compared to
the typical DF as seen in a narrow band image. 
Doppler measurements are usually strongly affected by the point
spread function (PSF) since it contaminates the resolved spectra
with other spectra originating from outside the position of the slit.
In the present case the PSF is actually making the Doppler measurement 
more robust. The fact that the high Doppler shifts are seen 
as a separate component results in a small effect from the PSF on these Doppler
measurements. This is because at the wavelength of the separate component 
other spectra (from neighboring spatial pixels)
usually are in the far wing. The low gradient of the far wing
results in a small effect on the Doppler component. Since this effect also
works vice versa the high Doppler shifts will show in spectra where
it otherwise would not, hence increasing the chances of observing the 
whole DF.

\section{Numerical experiments}\label{Sim}

To explain the lower velocities and decelerations in the present observations,
realistic simulations are needed.
The simulation presented by \citet{2006Hansteen} reproduce 
some of the key results from their observations, and it should therefore be 
well suited for our purpose. 
We solve the radiative transfer equations for the atmospheres given by 
these simulations. This is done by using the MULTI code \citep{MULTI}. 
The simulated atmospheres are given in two dimensions, but the
problem is simplified by solving for each column independently, 
each resolving $32.5$~km of the solar surface. Each column is solved 
in full NLTE for the Ca line, while we include the line blend in LTE,
but with a scattering term included in the source-function.
When solving the radiative transfer CRD is assumed; the results will not be 
significantly altered by PRD \citep{1989Uitenbroek}.

\begin{figure}[!ht]
\includegraphics[width=0.48\textwidth]{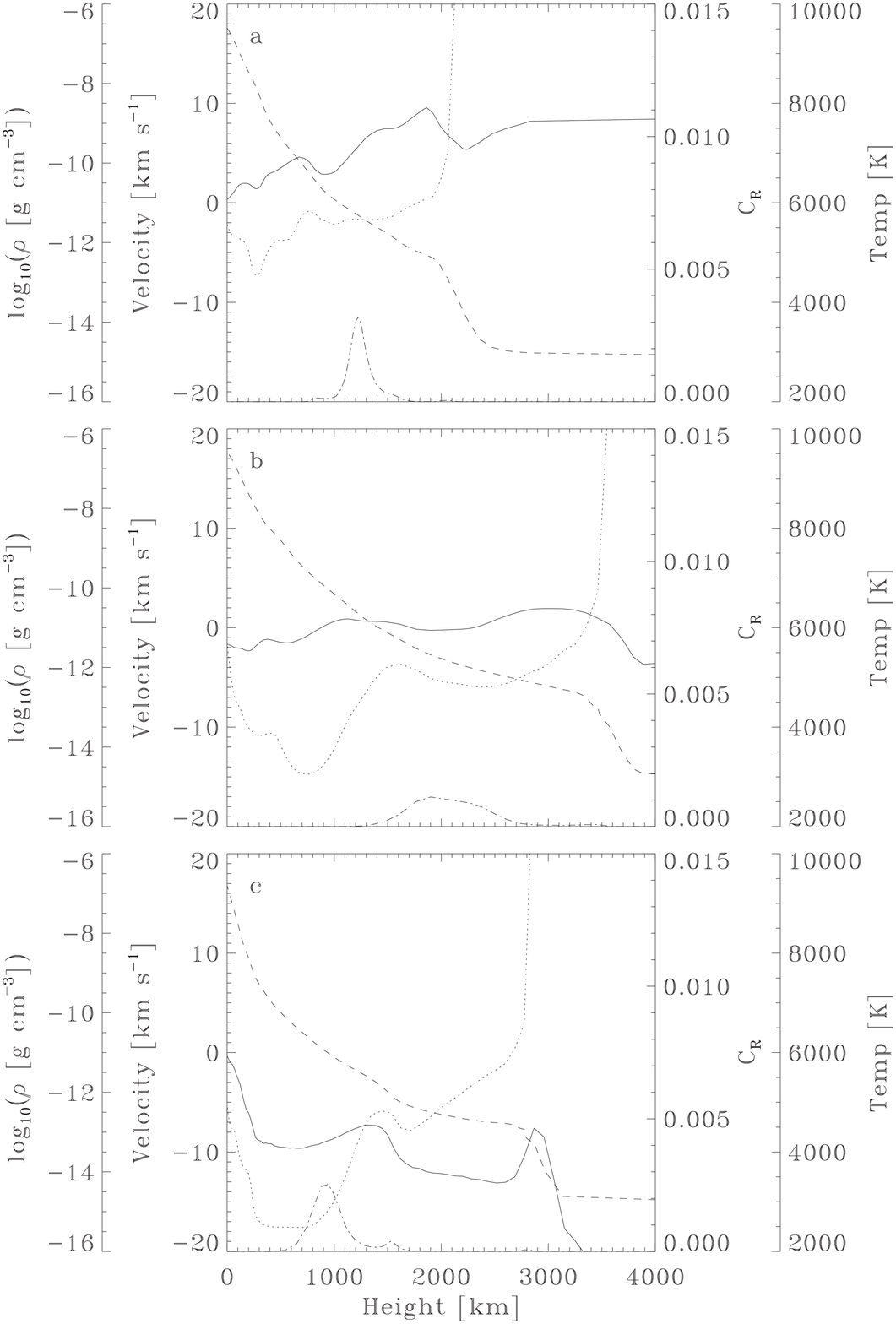}
\caption{
Atmospheric parameters for three different times in the life of a DF, 
20 seconds (panel a), 
100 seconds (panel b) and 160 seconds (panel c) are
seen. Velocity (solid line), temperature (dotted line),
logarithm of the density (dashed line), and finally the
contribution functions to relative intensity (dashed--dotted line)
at the Doppler shifted wavelength are plotted. 
}
\label{plotfive}
\end{figure}

Since the simulations have several drawbacks, such as instantaneous
hydrogen ionization, two dimensional geometry, and lack of global 
5--minute oscillations, it can not reproduce all details of the DFs.
The simulations can nevertheless be used to illustrate the general 
physics involved. For this purpose we have chosen a DF--like feature
in the simulations, which is analyzed in detail.
The DF--like feature is identified in a region in the simulations
where we have magnetic fields emerging from the photosphere.
In this region the fluxtubes are fairly vertical,
which is similar to the expected inclinations in the observations.
Several DF--like features can be identified in this region of the
simulations, and most of them show similar behavior as the one we 
investigate in detail.

\citet{2006Hansteen} and \citet{2007dePontieu} measure the movement of a 
bright top, which is caused by the steep temperature increase
in the transition region where it is believed that H$\alpha$ has some 
contribution. Hence they measure the movement of a point in the 
image plane, that is well defined due to the big contrast seen between 
the top of the DF and the surroundings. This observational 
method is compared with the spectroscopic method.

Since the calculation of the response function  to velocity (RFV) is 
based on linear approximations it is not possible to calculated RFV
for DFs, which are inherently non-linear in nature.
It is, however, possible to calculate the contribution functions to 
relative absorption (CFRA) \citep{1986Magain}. The CFRA are fairly similar 
to the RFV since the Doppler shifts are seen as an absorption component.
We calculate the contribution functions at the Doppler shifted wavelengths 
as measured in the different parts of the DF. The resulting contribution 
functions and the atmospheric velocity, temperature and density can be seen 
in Fig.\ref{plotfive}. 

The wave seen in panel A in Fig.\ref{plotfive} steepens as 
the atmospheric density drops about 4 magnitudes from the 
photosphere into the chromosphere. In the first stage of this DFs lifetime
the transition region is found at about $2$ Mm.
The Ca IR line samples a region below the
maximal velocity, which is seen close to the transition region.
At this time in the DF the measured Doppler shift is about $6$ km\,s${}^{-1}$,
while the velocity close to the transition region is about $10$ km\,s${}^{-1}$.
This difference in velocities is very similar to the differences between
the observations based on the motion of the TR and those based on the
Doppler shifts, see Fig.\ref{plotfour}.

In panel B in Fig.\ref{plotfive} the atmospheric parameters 
at a later time in this DFs lifetime are plotted. The transition region has
been pushed upwards by the shock, and it is now at a height of about $3.6$ Mm.
Even though the formation height of the Ca IR line is somewhat higher
in the atmosphere at this stage of the DF lifetime, it is still formed 
below the transition region. The formation height of the Ca IR line
is basically governed by the integrated density above the formation height,
hence the reason for the higher formation height at this stage of the DFs 
life time is the higher densities at equal geometric heights.
The extension in height of the CFRA is of some significance
to the measured Doppler shift, which in this case is close to zero. 

In panel C in Fig.\ref{plotfive} the atmospheric parameters at the end of
the DFs lifetime is plotted. The transition region height has  
declined, and is now seen at about $2.8$ Mm. The atmospheric velocity close
to the transition region is about $-13$ km\,s${}^{-1}$, while the measured
Doppler velocity is about $-8$ km\,s${}^{-1}$.
It should be noted that due to the lower height of formation in the
Ca~IR line the shocks sometimes overlap more than what is seen in the 
transition region. This leads to sometimes shorter lifetimes in the
DFs seen in the Doppler grams compared to the transition region.
Furthermore some of the movement of the transition region is apparent 
motion due to sideways motion of the DFs across the slit.

From the analysis above the intrinsic differences in the observational
methods can
explain the observed differences in maximal velocities. The difference in
deceleration can be explained in the shock framework, since the maximal 
velocities are lower, the deceleration must also be lower assuming similar
lifetimes (as mentioned above the lifetimes are somewhat shorter in the 
spectra, but the relative reduction in lifetimes is smaller then the relative
reduction in maximal velocities).

The DFs in the simulated spectra have lower maximal Doppler velocities,
than in the observations. 
Typical maximal values in the simulations are $9$ km\,s${}^{-1}$ versus 
$11$ km\,s${}^{-1}$ in the observations. 
Furthermore the spread in the maximal velocities 
in the simulations are lower than in the observations. These differences are
in part caused by a central reversal, which makes it impossible 
to measure the extreme Doppler shifts. The results are lower maximal 
velocities and life times, but no change in decelerations. 
These central reversals are caused  by the simplified two dimensional
geometry, which neglects the horizontal energy dissipation, hence increasing 
the temperature in the DFs. 
Tests from a three dimensional simulation show that this can account for 
some of the deficit in velocity, but not all. Other reasons can be 
differences in inclination angle of the local magnetic field in the 
observations and simulations. The shorter life times in the simulated DFs,
due to lack of global 5-minutes oscillations in the simulations 
will also affect the maximal velocities.
 
\section{Conclusions}\label{Con}

In this work we have presented co-spatial and co-temporal narrow band 
H$\alpha$ images and spectroscopic measurements of the 
Ca~{\small{II}}~8662~{\AA} line. These observations have been 
used to identify 26 DFs, and measure their Doppler shifts. 
A reduced--$\chi^2$ analysis shows that the time evolution of the
Doppler shifts are  
well approximated by a linear fit, if the measurement errors are about 
$2$~km\,s${}^{-1}$. Using this approximation we derive values for
the decelerations and maximal velocities for each DF. 
Scatter plots of the deceleration and maximal velocity show a strong 
positive correlation between the two. We also observe weak correlations 
between the deceleration and lifetime and the maximal velocity and lifetime. 
These results are supporting the shock-wave theory as explanation 
model for the DFs \citep{2006Hansteen, 2007dePontieu,2007Lars}.

Furthermore, the Doppler shifts show that at least a subset of DFs are
caused by mass moving up and down in the atmosphere. 
The values of the maximum velocity and 
decelerations are all somewhat lower than earlier reported values. 
Using numerical experiments we have explained the differences
in the two observational sets with the intrinsic 
differences in observational methods.
Earlier observations have used the high contrast seen between the
top of the DF and the background for measuring the proper motion of the DF.
This high contrast is caused by the intensity increase due to contributions
from the transition region.
In the present observations the DF motion is measured using Doppler shifts,
which are affected by the atmospheric conditions over the formation height. 
The formation height of the Ca~{\small{II}}~8662~{\AA}~line
is much lower than the transition 
region. Since the shock amplitude is increasing from the 
formation height for the Ca IR line to the transition region we
necessarily measure lower velocities using spectroscopy.
The difference in maximal velocities derived from the two methods in our 
simulations is about a factor two, which is about the same as observed.

The one dimensional nature of the slit spectrograph
somewhat affects our results, but experiments with narrow band images
show that this is not altering the results significantly.

\acknowledgements
We thank Peter S{\"u}tterlin for helping with the 
acquisition and reduction of the DOT data and
Marte Skogvoll for general help with the observations. 
{\O}L thank Margrethe Wold for discussions about 
the statistical method.
This research was supported by the European Community's Human Potential
Program through the TOSTISP (contract HPRN-CT-2002-00310) program and by The 
Research Council of Norway through grant 170935/v30 and through
grants of computing time from the Programme for Supercomputing.
The Swedish 1-m Solar Telescope is operated on the island of La Palma
by the Institute for Solar Physics of the Royal Swedish Academy of
Sciences in the Spanish Observatorio del Roque de los Muchachos of the
Instituto de Astrof{\'\i}sica de Canarias.
The Dutch Open Telescope, operated by Utrecht University at the 
Spanish Observatorio del Roque de los Muchachos of the 
Instituto de Astrof{\'\i}sica de Canarias.
This research has made use of NASA's Astrophysics Data System.
\bibliographystyle{apj}
\bibliography{paper2}

\begin{thebibliography}{37}
\expandafter\ifx\csname natexlab\endcsname\relax\def\natexlab#1{#1}\fi

\bibitem[{{Allende Prieto} \& {Garcia Lopez}(1998)}]{1998Allende}
{Allende Prieto}, C., \& {Garcia Lopez}, R.~J. 1998, \aaps, 129, 41

\bibitem[{{Beckers}(1968)}]{1968Beckers}
{Beckers}, J.~M. 1968, \solphys, 3, 367

\bibitem[{{Bel} \& {Leroy}(1977)}]{1977Bel}
{Bel}, N., \& {Leroy}, B. 1977, \aap, 55, 239

\bibitem[{{Brault} \& {Neckel}(1987)}]{1987Brault}
{Brault}, J., \& {Neckel}, H. 1987, Hamburg Observatory anonymous ftp:
  \texttt{ftp.hs.uni-hamburg.de}

\bibitem[{Carlsson(1986)}]{MULTI}
Carlsson, M. 1986, A Computer Program for Solving Multi-Level Non-LTE Radiative
  Transfer Problems in Moving or Static Atmospheres (Uppsala Astronomical
  Observatory: Report No.\ 33)

\bibitem[{{Christopoulou} {et~al.}(2001){Christopoulou}, {Georgakilas}, \&
  {Koutchmy}}]{2001Christo}
{Christopoulou}, E.~B., {Georgakilas}, A.~A., \& {Koutchmy}, S. 2001, \solphys,
  199, 61

\bibitem[{{De Pontieu} {et~al.}(2005){De Pontieu}, {Erd{\'e}lyi}, \& {De
  Moortel}}]{2005dePontieu}
{De Pontieu}, B., {Erd{\'e}lyi}, R., \& {De Moortel}, I. 2005, \apjl, 624, L61

\bibitem[{{De Pontieu} {et~al.}(2003){De Pontieu}, {Erd{\'e}lyi}, \& {de
  Wijn}}]{2003dePontieu}
{De Pontieu}, B., {Erd{\'e}lyi}, R., \& {de Wijn}, A.~G. 2003, \apjl, 595, L63

\bibitem[{{De Pontieu} {et~al.}(2004){De Pontieu}, {Erd{\'e}lyi}, \&
  {James}}]{2004dePontieu}
{De Pontieu}, B., {Erd{\'e}lyi}, R., \& {James}, S.~P. 2004, \nat, 430, 536

\bibitem[{{De Pontieu} {et~al.}(2007){De Pontieu}, {Hansteen}, {Rouppe van der
  Voort}, {van Noort}, \& {Carlsson}}]{2007dePontieu}
{De Pontieu}, B., {Hansteen}, V.~H., {Rouppe van der Voort}, L., {van Noort},
  M., \& {Carlsson}, M. 2007, \apj, 655, 624

\bibitem[{{de Wijn} \& {de Pontieu}(2006)}]{2006deWijn}
{de Wijn}, A.~G., \& {de Pontieu}, B. 2006, \aap, 460, 309

\bibitem[{{Hansteen} {et~al.}(2006){Hansteen}, {De Pontieu}, {Rouppe van der
  Voort}, {van Noort}, \& {Carlsson}}]{2006Hansteen}
{Hansteen}, V., {De Pontieu}, B., {Rouppe van der Voort}, L., {van Noort}, M.,
  \& {Carlsson}, M. 2006, \apjl, 647, L73

\bibitem[{{Heggland} {et~al.}(2007){Heggland}, {De Pontieu}, \&
  {Hansteen}}]{2007Lars}
{Heggland}, L., {De Pontieu}, B., \& {Hansteen}, V.~H. 2007, \apj, 666, 1277

\bibitem[{{Koza} {et~al.}(2007){Koza}, {S\"utterlin}, {Ku\v cera}, \&
  {Ryb\'ak}}]{2007Julius}
{Koza}, J., {S\"utterlin}, P., {Ku\v cera}, A., \& {Ryb\'ak}, J. 2007, in ASP
  Conf. Ser. 368, ed. P.~{Heinzel}, I.~{Dorotovic}, \& R.~{Rutten}

\bibitem[{{Kupka} {et~al.}(1999){Kupka}, {Piskunov}, {Ryabchikova}, {Stempels},
  \& {Weiss}}]{vald}
{Kupka}, F., {Piskunov}, N., {Ryabchikova}, T.~A., {Stempels}, H.~C., \&
  {Weiss}, W.~W. 1999, \aaps, 138, 119

\bibitem[{{Langangen} {et~al.}(2007){Langangen}, {Carlsson}, \& {Rouppe van der
  Voort}}]{2007Langangen}
{Langangen}, {\O}., {Carlsson}, M., \& {Rouppe van der Voort}, L. 2007, \apj,
  655, 615

\bibitem[{{Magain}(1986)}]{1986Magain}
{Magain}, P. 1986, \aap, 163, 135

\bibitem[{{Michalitsanos}(1973)}]{1973Mich}
{Michalitsanos}, A.~G. 1973, \solphys, 30, 47

\bibitem[{{Nishikawa}(1988)}]{1988Nishikawa}
{Nishikawa}, T. 1988, \pasj, 40, 613

\bibitem[{{Piskunov} {et~al.}(1995){Piskunov}, {Kupka}, {Ryabchikova}, {Weiss},
  \& {Jeffery}}]{1995Piskunov}
{Piskunov}, N.~E., {Kupka}, F., {Ryabchikova}, T.~A., {Weiss}, W.~W., \&
  {Jeffery}, C.~S. 1995, \aaps, 112, 525

\bibitem[{{Press} {et~al.}(1992){Press}, {Teukolsky}, {Vetterling}, \&
  {Flannery}}]{Numerical}
{Press}, W.~H., {Teukolsky}, S.~A., {Vetterling}, W.~T., \& {Flannery}, B.~P.
  1992, {Numerical recipes in FORTRAN. The art of scientific computing}
  (Cambridge: University Press, |c1992, 2nd ed.)

\bibitem[{{Rimmele}(2000)}]{2000Rimmele}
{Rimmele}, T.~R. 2000, in Proc. SPIE Vol. 4007, p. 218-231, Adaptive Optical
  Systems Technology, Peter L. Wizinowich; Ed., ed. P.~L. {Wizinowich},
  218--231

\bibitem[{{Rouppe van der Voort} {et~al.}(2007){Rouppe van der Voort}, {De
  Pontieu}, {Hansteen}, {Carlsson}, \& {van Noort}}]{2007Rouppe}
{Rouppe van der Voort}, L.~H.~M., {De Pontieu}, B., {Hansteen}, V.~H.,
  {Carlsson}, M., \& {van Noort}, M. 2007, \apjl, 660, L169

\bibitem[{{Rutten} {et~al.}(2004){Rutten}, {Hammerschlag}, {Bettonvil},
  {S{\"u}tterlin}, \& {de Wijn}}]{DOT}
{Rutten}, R., {Hammerschlag}, R., {Bettonvil}, F., {S{\"u}tterlin}, P., \& {de
  Wijn}, A. 2004, \aap, 413, 1183

\bibitem[{{Ryabchikova} {et~al.}(1999){Ryabchikova}, {Piskunov}, {Stempels},
  {Kupka}, \& {Weiss}}]{1999Ryabchikova}
{Ryabchikova}, T., {Piskunov}, N., {Stempels}, H., {Kupka}, F., \& {Weiss}, W.
  1999, Physica Scripta, T83, 162

\bibitem[{{Scharmer} {et~al.}(2003{\natexlab{a}}){Scharmer}, {Bjelksj{\"o}},
  {Korhonen}, {Lindberg}, \& {Petterson}}]{SST}
{Scharmer}, G., {Bjelksj{\"o}}, K., {Korhonen}, T., {Lindberg}, B., \&
  {Petterson}, B. 2003{\natexlab{a}}, in Proceedings of the SPIE, Volume 4853.
  pp. 341-350

\bibitem[{{Scharmer} {et~al.}(2003{\natexlab{b}}){Scharmer}, {Dettori},
  {L{\"o}fdahl}, \& {Shand}}]{SSTAO}
{Scharmer}, G.~B., {Dettori}, P.~M., {L{\"o}fdahl}, M.~G., \& {Shand}, M.
  2003{\natexlab{b}}, in Innovative Telescopes and Instrumentation for Solar
  Astrophysics. Edited by Stephen L. Keil, Sergey V. Avakyan . Proceedings of
  the SPIE, Volume 4853, pp. 370-380

\bibitem[{{Shine} {et~al.}(1994){Shine}, {Title}, {Tarbell}, {Smith}, {Frank},
  \& {Scharmer}}]{1994DS}
{Shine}, R.~A., {Title}, A.~M., {Tarbell}, T.~D., {Smith}, K., {Frank}, Z.~A.,
  \& {Scharmer}, G. 1994, \apj, 430, 413

\bibitem[{{Suematsu}(1990)}]{1990Suematsu}
{Suematsu}, Y. 1990, in LNP Vol. 367: Progress of Seismology of the Sun and
  Stars, ed. Y.~{Osaki} \& H.~{Shibahashi}, 211

\bibitem[{{Suematsu} {et~al.}(1995){Suematsu}, {Wang}, \&
  {Zirin}}]{1995Suematsu}
{Suematsu}, Y., {Wang}, H., \& {Zirin}, H. 1995, \apj, 450, 411

\bibitem[{{Tsiropoula} {et~al.}(1994){Tsiropoula}, {Alissandrakis}, \&
  {Schmieder}}]{1994Tsiropoula}
{Tsiropoula}, G., {Alissandrakis}, C.~E., \& {Schmieder}, B. 1994, \aap, 290,
  285

\bibitem[{{Tziotziou} {et~al.}(2004){Tziotziou}, {Tsiropoula}, \&
  {Mein}}]{2004Kostas}
{Tziotziou}, K., {Tsiropoula}, G., \& {Mein}, P. 2004, \aap, 423, 1133

\bibitem[{{Uitenbroek}(1989)}]{1989Uitenbroek}
{Uitenbroek}, H. 1989, \aap, 213, 360

\bibitem[{{Uitenbroek}(2006)}]{2006Uitenbroek}
---. 2006, \apj, 639, 516

\bibitem[{{van Noort} {et~al.}(2005){van Noort}, {Rouppe van der Voort}, \&
  {L{\"o}fdahl}}]{momfbd}
{van Noort}, M., {Rouppe van der Voort}, L., \& {L{\"o}fdahl}, M.~G. 2005,
  \solphys, 228, 191

\bibitem[{{van Noort} \& {Rouppe van der Voort}(2006)}]{2006Noort}
{van Noort}, M.~J., \& {Rouppe van der Voort}, L.~H.~M. 2006, \apjl, 648, L67

\bibitem[{{von der Luehe}(1993)}]{speckle}
{von der Luehe}, O. 1993, \aap, 268, 374

\end{thebibliography}

\end{document}